\documentclass[twocolumn,a4paper,amsmath,amssymb,showpacs,longbibliography,jpsj,superscriptaddress]{revtex4-1}
\usepackage{graphicx}
\usepackage{color}
\usepackage{bm}
\usepackage[normalem]{ulem}
\usepackage{braket}
\usepackage{caption}

\begin{document}

\title{Wannier Based Analysis of the Direct-Indirect Bandgap Transition by Stacking MoS$_2$ Layers}
\author{Shunsuke Hirai}
\thanks{Contact author: ss25567q@st.omu.ac.jp}

\affiliation{Department of Materials Science, Graduate School of Engineering, Osaka Metropolitan University, Sakai, Osaka 599-8531, Japan}
\author{Ibuki Terada}
\affiliation{Department of Materials Science, Graduate School of Engineering, Osaka Metropolitan University, Sakai, Osaka 599-8531, Japan}
\author{Michi-To Suzuki}
\thanks{Corresponding author: mts@omu.ac.jp}
\affiliation{Department of Materials Science, Graduate School of Engineering, Osaka Metropolitan University, Sakai, Osaka 599-8531, Japan}
\affiliation{Center for Spintronics Research Network, Graduate School of Engineering Science, Osaka University, Toyonaka, Osaka 560-8531, Japan}

\date{April 2026}

\begin{abstract}
Molybdenum disulfide (MoS$_2$), a layered van der Waals material, has attracted considerable attention as a promising alternative to graphene for applications in field-effect transistors and nanophotonic devices because of its sizable band gap, high carrier mobility, large on/off ratio, and strong photoluminescence efficiency. A particularly intriguing property of MoS$_2$ is the transition of its band gap character with layer thickness: while the monolayer exhibits a direct gap, the band gap becomes indirect in multilayer and bulk forms.In this study, we clarify the microscopic mechanism underlying this transition. Focusing on the roles of atomic orbitals and interlayer interactions, we perform an analysis combining first-principles calculations with a Wannier-based model. Although interlayer $p_z$--$p_z$ coupling between neighboring sulfur atoms has been recognized as a key factor in this transition, we find that a complete quantitative description additionally requires interlayer $p_z$--$p_x$ and $p_z$--$p_y$ couplings between neighboring sulfur atoms. These findings highlight the importance of both out-of-plane and in-plane orbital contributions in governing the electronic structure of layered MoS$_2$, providing deeper insight into its band gap engineering for future device applications.
\end{abstract}

\maketitle

\section{Introduction}
Transition metal dichalcogenides (TMDs) have emerged as a fascinating class of materials owing to their layered structures and the diverse physical properties that arise from reduced dimensionality\cite{chhowalla2015two,liu2015electronic,kuc2015electronic}. Because the inter layer bonding is relatively weak, dominated by van der Waals interactions, individual monolayers can be readily exfoliated, making TMDs ideal platforms for exploring novel two-dimensional phenomena. Their thin-film forms often exhibit electronic, optical, and mechanical properties that differ markedly from those of the bulk, which has spurred intensive interest in their potential applications in nanoelectronics or optoelectronics.

Among TMDs, molybdenum disulfide (MoS$_2$) has received particular attention due to its remarkable electronic properties~\cite{LI2015,YAZYEV2015}. One of the most striking properties of MoS$_2$ is the change in bandgap character with layer thickness: bulk MoS$_2$ is an indirect-gap semiconductor, whereas monolayer MoS$_2$ exhibits a direct gap. 
This behavior was first predicted theoretically~\cite{Li2007} and later confirmed experimentally through photoluminescence measurements~\cite{Splendiani2010,MaK2010,Eda2011}. These pioneering studies triggered extensive research efforts into the fundamental physics of monolayer MoS$_2$ and related TMDs.

From an application standpoint, the unique layer-dependent band structure of MoS$_2$ makes it a strong candidate for flexible semiconductors, optoelectronic devices, and field-effect transistors (FETs). In particular, its high carrier mobility and excellent on/off current ratio have made MoS$_2$-based FETs a central focus of device-oriented studies~\cite{Radisavljevic2011}. For such applications, understanding and controlling the evolution of the electronic structure with layer number is of crucial importance. Therefore, considerable effort has been devoted to achieving a deeper understanding of this phenomenon. In particular, because tight-binding (TB) models are widely used in the study of MoS$_2$-based devices and applications, accurately capturing this phenomenon is essential for constructing reliable TB models~\cite{Cappelluti2013}. Although various TB models based on different basis sets have been proposed ~\cite{zahid2013,liu2013,Rostami2013,ridolfi2015,Cappelluti2013,Fang2015}, it is well known that accurate reproduction of the valence and conduction bands near the Fermi level requires inclusion of the Mo 4$d$ orbitals and the S 3$p$ orbitals. However, multilayer TB models of MoS$_2$ that incorporate interlayer interactions still face challenges in reproducing parts of the conduction band, and further improvement is required. In this study, based on a Wannier-based bulk TB model, we appropriately treated the interlayer interactions and were able to accurately reproduce the band structures of multilayer systems. Using this TB model and DFT calculations, we provide a detailed analysis of the microscopic mechanism behind the direct-to-indirect band gap transition in MoS$_2$ induced by increasing layers. We clarify the roles of specific atomic orbitals and interlayer interactions in shaping the electronic band structure. Our results not only deepen the fundamental understanding of layer-dependent band structure evolution but also provide insights into strategies for band gap engineering in van der Waals materials for future device applications.

\section{Calculation Details}\label{calmethod}
\begin{figure}
\includegraphics[width=0.8\linewidth]{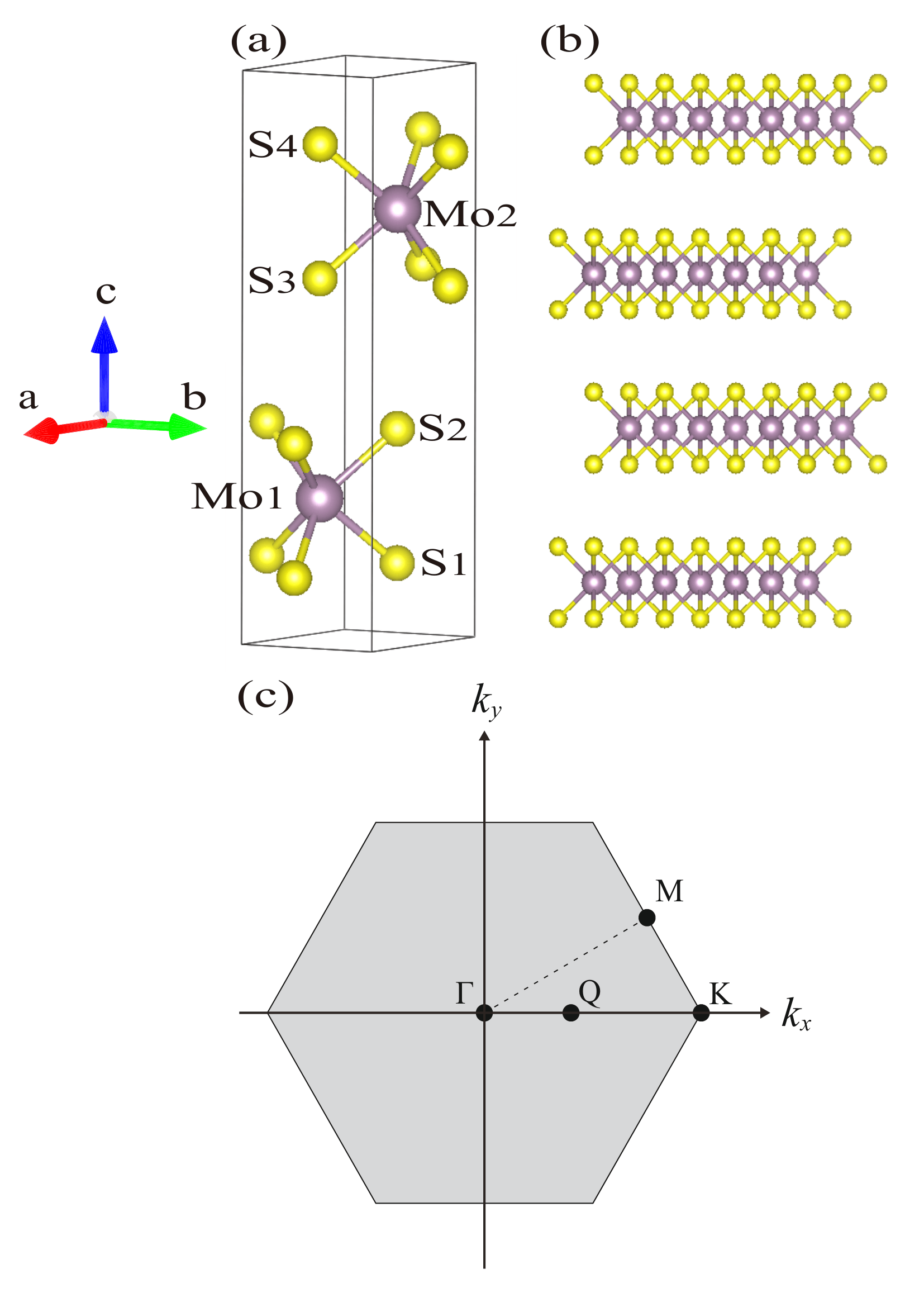}
\caption{Crystal structure of 2H-MoS$_2$: (a) unit cell and (b) side view of layer structure. These figures are drawn by VESTA\cite{VESTA}. (c) Brillouin Zone of MoS$_2$}
\label{Fig:crystal}
\end{figure}

 The crystal structure of 2H-MoS$_2$ is shown in Fig.~\ref{Fig:crystal}. 
 The hexagonal unit cell contains two Mo atoms and four S atoms. Each Mo atom is coordinated by six S atoms in a trigonal prismatic arrangement, forming S–Mo–S layers stacked along the c-axis in an ABA sequence with weak van der Waals interactions between adjacent layers. The space group is P63/mmc with 2c sites, $(\frac{1}{3},\frac{2}{3},\frac{1}{4})$ and $(\frac{2}{3},\frac{1}{3},\frac{3}{4})$, for Mo and 4f sites, $(\frac{1}{3},\frac{2}{3},u)$ $(\frac{2}{3},\frac{1}{3},\bar{u})$ $(\frac{1}{3},\frac{2}{3},\frac{1}{2}-u)$ and $(\frac{2}{3},\frac{1}{3},\frac{1}{2}+u)$, for S. The experimentally determined lattice constants are $a=3.160$~\AA ~and $c=12.294$~\AA, and the internal parameter is $u=0.621$~\cite{Pauling1923}.
 
 In this study, first-principles calculations were performed within Density Functional Theory (DFT) using generalized gradient approximation (GGA) with the PBE sol functional, implemented in \textsc{Quantum ESPRESSO} ~\cite{QE1,QE2}. The plane-wave cutoff was set to 150~Ry for the wavefunctions and 600~Ry for the charge density. For the bulk system, an $8 \times 8 \times 8$ Monkhorst--Pack $k$-point mesh was used for self consistent field (SCF) calculations and an $18 \times 18 \times 18$ mesh for the calculations of Density of States. The optimized lattice constants in bulk form were $a=3.14$\, \text{\AA}\, and $c=12.6$\, \AA, with the internal parameter determined as ${u} = 0.6259$. For the monolayer system, a supercell approach was employed with a vacuum spacing of about 60 \text{\AA}\, between periodic images. An $8 \times 8 \times 2$ $k$-point mesh was adopted for the monolayer supercell SCF calculation. A Tight-binding model was constructed using \textsc{Wannier90} ~\cite{WANNIER} based on the bulk electronic structure calculated with \textsc{Quantum ESPRESSO}. 

In this study, we choose the Mo-4$d$ and S-3$p$ orbitals as projection orbitals that primarily contribute to the electronic states near the Fermi level~\cite{ridolfi2015,Fang2015,Cappelluti2013}.

To obtain multi-layer electronic states, we constructed 2D $\boldsymbol{k}$-space Hamiltonian for an multilayer system using the bulk tight-binding (TB) model.
We obtained the hopping integrals $t^{ij}(\boldsymbol{R}-\boldsymbol{R}^{\prime})$ from the TB model, where $\bm R$ is the lattice vector and the index $i=(A,\alpha)$ denotes a set of atomic species $A$ and orbitals $\alpha$ in the unit cell.
Using the hopping integrals $t^{ij}(\boldsymbol{R}-\boldsymbol{R}^{\prime})$, we can freely construct the $\boldsymbol{k}$-space Hamiltonian for the $2N$-layer system.
We here impose the open boundary conditions along the $c$-axis and the periodic boundary conditions along the $a$- and $b$-axes to describe the multilayer system. Under this setup, the $\boldsymbol{k}$-space Hamiltonian for a $2N$-layer system is written as follows:
\begin{align}
\label{Hamiltonian}
&\mathcal H_{\boldsymbol{k}}^{2N} = \sum_{i,j} \sum_{R_z,R^\prime_z=1}^N\ket{iR_z\boldsymbol{k}}\mathcal H_{\boldsymbol{k}}^{ij}(R_z-R^\prime_z)\bra{jR^\prime_z\boldsymbol{k}},\\
\label{Hamiltonian2}
&\mathcal{H}_{\boldsymbol{k}}^{ij}(R_z-R^{\prime}_z)= \sum_{R_x,R_y}\sum_{R^\prime_x,R^\prime_y} t^{ij}(\boldsymbol{R}-\boldsymbol{R^\prime})e^{-i\boldsymbol{k}\cdot(\boldsymbol{R}-\boldsymbol{R^\prime})}.
\end{align}
Here, $\ket{iR_z\bm k}\bra{iR_z\bm k}$ represents the projection operator onto the atomic orbital $\alpha$ of atom $A$~(labeled by $i=(A,\alpha)$) in the unit cell at $R_z$ for the wavenumber $\bm k=(k_x,k_y)$. The Hamiltonian for the $(2N-1)$-layer system is obtained by excluding the top layer in the unit cell at $R_z=N$ from the $2N$-layer Hamiltonian.

\section{Results and Discussion}
In this study, we investigated the mechanism behind the transition of the energy-gap structure of MoS$_2$ from a direct transition in the monolayer to an indirect transition in the multilayer structure based on first-principles calculations.
To investigate in detail the relationship between the electronic states and atomic orbitals, we also constructed a TB model based on first-principles calculations as mentioned in section \ref{calmethod}.

\begin{figure}[t]
\includegraphics[width=0.95\linewidth]{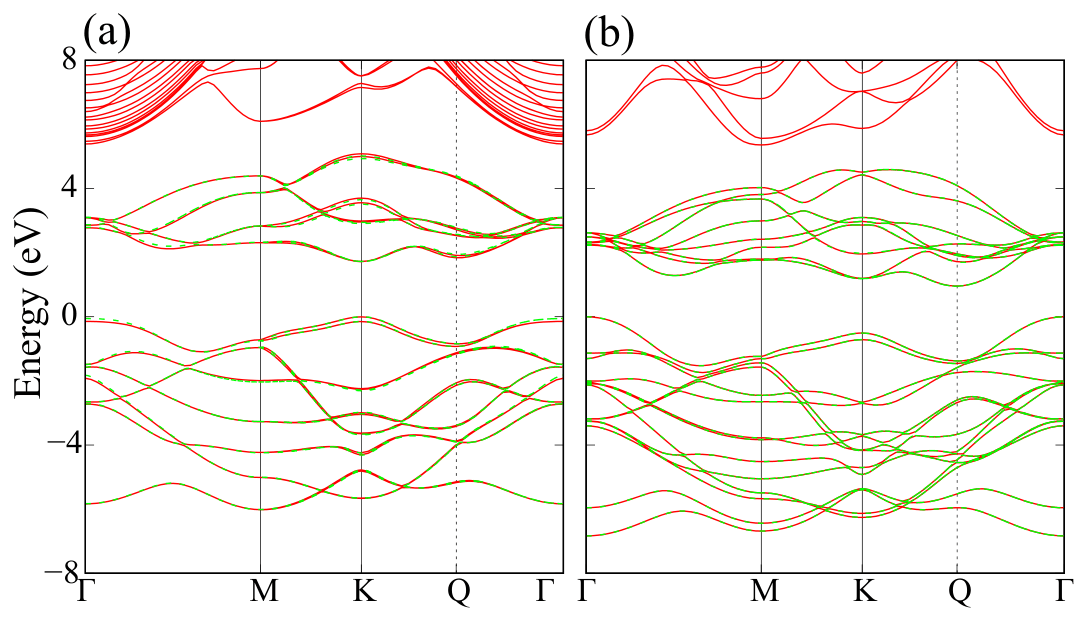}
\caption{Comparison of band structure by DFT(red solid lines) and TB model(green broken lines) : (a) bulk and (b) monolayer MoS$_2$. The arrows indicate the band-gap transitions. }
\label{Fig:QEWANNIERband}
\end{figure}

Our TB model for the bulk MoS$_2$ with WANNIER90 well reproduces the DFT calculation, as shown in Fig.~\ref{Fig:QEWANNIERband} (a). Based on this bulk TB model, we constructed the multilayer Hamiltonian using Eq.~\eqref{Hamiltonian}. The band structure of monolayer MoS$_2$ calculated from Eq. ~\eqref{Hamiltonian} is shown by the green dashed line in Fig.~\ref{Fig:QEWANNIERband} (b), and its good agreement with the DFT results validates the applicability of our TB model to MoS$_2$ systems with arbitrary layer numbers.

Here, we revisit the direct-indirect bandgap transition in MoS$_2$. The electronic transition in the monolayer case occurs between valence and conduction bands at the $K$ point. In contrast, in the bulk case, the electronic transition occurs between the valence-band maximum at the $\Gamma$ and the conduction-band minimum at the $Q$, where $Q$ lies on the $K$--$\Gamma$ path, as shown in Fig.~\ref{Fig:crystal}(c). The band gap estimated by our calculations respectively are $1.72$ eV in the monolayer and $0.958$ eV in the bulk, which is consistent with previous studies~\cite{Li2007,Splendiani2010,cheiw2012,EugeneS2012,ahmad2014comparative}.

\begin{figure}[t]
\includegraphics[width=0.8\linewidth]{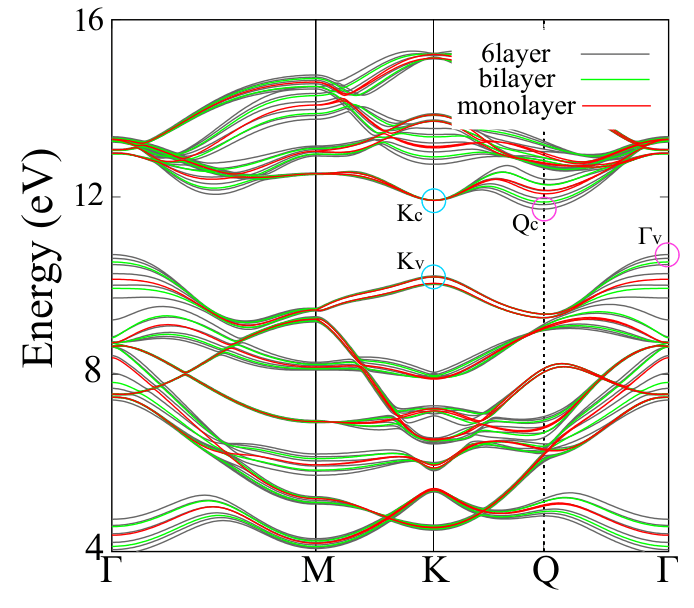}
\caption{Band structure for monolayer (green), bilayer (red), and 6-layers (gray) MoS$_2$. 
}
\label{Fig:WANNIERlayerband}
\end{figure}

Several studies have suggested that interlayer interactions between neighboring S atoms are important for the direct--indirect bandgap transition, and that among them the $p_z$ orbital is particularly important because of its orientation perpendicular to the layers~\cite{Cappelluti2013,Fang2015}. Fang et al. proposed a TB model in which the interlayer S--S interaction for atomic pairs separated by 5~\AA\ or less is incorporated as an interlayer interaction term and which also includes the interlayer coupling between the Mo $d_{z^2}$ and S $p_z$ orbitals~\cite{Fang2015}. However, this model does not fully capture the band structure around the $Q$ point in the conduction band of the bulk system, suggesting that a complete description of the direct--indirect bandgap transition requires quantitative evaluation of the interlayer interactions that affect the band splitting and their incorporation into the model.
To examine how the band structure evolves with increasing number of layers, we calculated the band structures for monolayer, bilayer and 6-layer systems obtained by Eq.~\eqref{Hamiltonian} in Fig.~\ref{Fig:WANNIERlayerband}. 
Significant modifications in the overall band structure arise from noticeable band splittings appearing in certain regions of the Brillouin zone. The band splitting is small at $K$ point indicated by the blue circles in Fig.~\ref{Fig:WANNIERlayerband}. Hereafter, these points are referred to as $K_\mathrm{v}$ for valance band top and $K_\mathrm{c}$ for conduction band bottom. On the other hand, the band splitting is large at the $Q$ and $\Gamma$ points indicated by the red circles ($Q_\mathrm{c}$ and $\Gamma_\mathrm{v}$). Consequently, the bottom of conduction-band or the top of valence-band move from the points with small band splitting to those where the splitting is large as the number of layers increases. Specifically, the valence-band top shifts from the $K$ point to the $\Gamma$ point, while the conduction-band bottom shifts from the $K$ point to the $Q$ point. Understanding what causes the band splitting is crucial for elucidating the mechanism of the direct-indirect bandgap transition of MoS$_2$. 

\begin{figure}[t]
    \centering
    \includegraphics[width=0.95\linewidth]{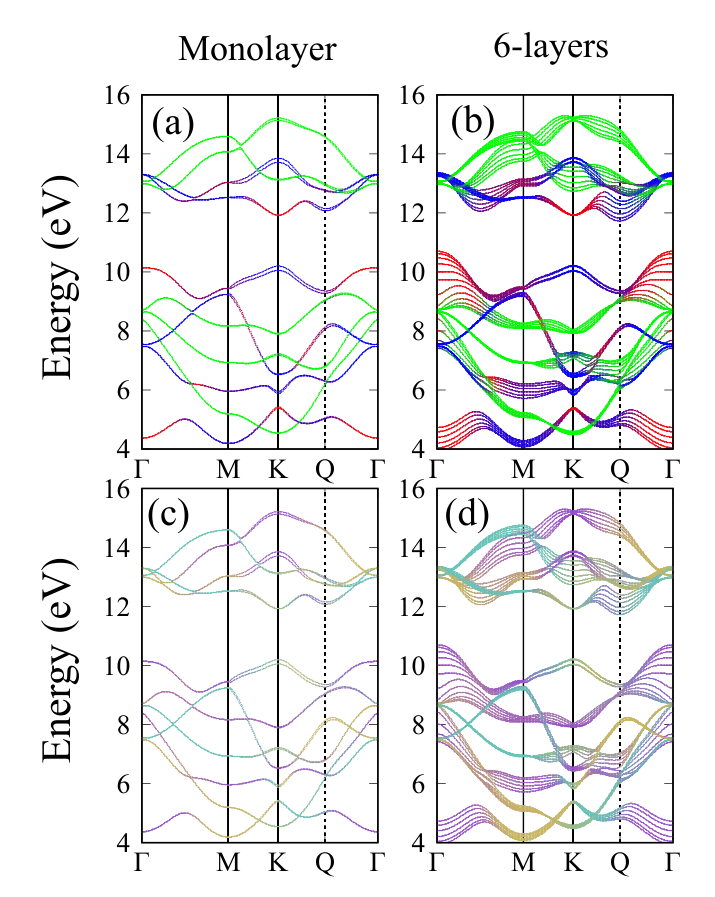}

    \caption{(a)(b) The Mo $d$ orbital character of the MoS$_2$ band structure for (a) monolayer and (b) 6-layer. The orbital weights are represented as follows: $d_{z^2}$ (red), $d_{xz}$ and $d_{yz}$ (green), and $d_{xy}$ and $d_{x^2-y^2}$ (blue). 
    (c)(d) The S $p$ orbital character of the MoS$_2$ band structure for (c) monolayer and (d) 6-layer. The orbital weights are represented as follows: $p_z$ (purple), $p_x$ (orange), and $p_y$ (light blue).}
    \label{Fig:weightband}
\end{figure}

\begin{table}[t]
    \begin{minipage}{0.9\linewidth}
        \centering
        \captionof{table}{Orbital weights at the high-symmetry points and the Q point.}
        \begin{tabular}{c|c|c|c|c}
             & Mo $d_{z^2}$ & Mo $d_{xy}$ and $d_{x^2-y^2}$ & S $p_z$ & S $p_x$ and S $p_y$ \\
            \hline
             $\Gamma_\mathrm{v}$ & 58.5\% & -  & 41.2\% & -  \\
            $K_\mathrm{v}$ & -  & 82.0\% & - & 17.7\% \\
            $K_\mathrm{c}$ & 82.6\% & - & - & 17.2\% \\
            $Q_\mathrm{c}$ &10.7\% & 47.8\% & 12.1\% & 26.1\%
        \end{tabular}

        \label{tab:weight}
    \end{minipage}
\end{table}

We discuss the band splitting from the perspective of atomic orbitals, based on our multilayer TB model~\eqref{Hamiltonian}.
Figure~\ref{Fig:weightband} shows the contributions of the Mo $4d$ and the S $3p$ orbitals to the energy bands of the monolayer and the six-layers. We can identify the atomic orbitals responsible for the band splitting associated with the direct-indirect bandgap transition. Table~\ref{tab:weight} summarizes the orbital weights at the $\Gamma,~K$ and $Q$ points.

First, we discuss the band splitting of the valence band. The Mo $d_{z^2}$ orbital and S $p_z$ orbital are dominant at the $\Gamma_\mathrm{v}$ point where the bands splitting is large. In contrast, Mo $d_{x^2-y^2}$ and $d_{xy}$ orbitals and the S $p_x$ and $p_y$ orbitals are dominant at the $K_\mathrm{v}$ point where the band splitting is small. 
Our results suggest that the band splitting in the valence band originates from the interlayer coupling between out-of-plane orbitals. Therefore, the large band splitting at $\Gamma_{\mathrm v}$ is described by a following picture : S atoms are the closest atoms in adjacent layers, and the interlayer interaction between their out-of-plane $p_z$ orbitals is therefore particularly strong. Moreover, we calculated the matrix elements of $\boldsymbol{k}$-space Hamiltonian~\eqref{Hamiltonian2} and summarized its matrix elements corresponding interlayer couplings at $\Gamma$, $Q$ and  $K$ points, in Table~\ref{Table:Hamiltonian}. At the $\Gamma$ point, the interlayer coupling between S $p_z$ orbitals in neighboring layers are significantly larger than other interlayer matrix elements. On the other hand, at the $K$ points, interlayer interaction elements in Hamiltonian have a small absolute value. These results provide quantitative confirmation that S $p_z$ orbitals exhibit strong interlayer interactions governing the valence band splitting.

\begin{table*}[t]
     \caption{Matrix elements corresponding to the interlayer interaction in the Hamiltonian arranged in order of absolute value (eV) at each $\boldsymbol{k}$-point. The label is a pair of $(i,j)$ in Eq.(2). The atomic labels are illustrated in Fig.~\ref{Fig:crystal}.}
     \label{Table:Hamiltonian}
    \begin{tabular}{cc|cc|cc}
   $Q$ point  &  absolute value &  $K$ point  &  absolute value &  $\Gamma$ point  &  absolute value \\ \hline
   $(\text{S}_2p_z,\text{S}_3p_z)$  & 0.696 & $(\text{S}_2p_x,\text{S}_3p_x)$ & 0.105 & $(\text{S}_2p_z,\text{S}_3p_z)$ & 0.932 \\
   $(\text{S}_2p_z,\text{S}_3p_y)$  & 0.250 & $(\text{S}_2p_y,\text{S}_3p_y)$ & 0.104 & $(\text{Mo}_1d_{z^2},\text{S}_3p_z)$ & 0.161 \\
   $(\text{S}_2p_z,\text{S}_3p_x)$  & 0.159 & $(\text{S}_2p_x,\text{S}_3p_y)$ & 0.102 & $(\text{Mo}_1d_{z^2},\text{Mo}_2d_{z^2})$ & 0.072 \\
   $(\text{Mo}_1d_{x^2-y^2},\text{S}_3p_z)$  & 0.100 & $(\text{Mo}_1d_{z^2},\text{S}_3p_z)$ & 0.037 & $(\text{S}_1p_z,\text{S}_3p_z)$ & 0.046 \\
   $(\text{S}_2p_x,\text{S}_3p_x)$  & 0.099 & $(\text{Mo}_1d_{xy},\text{S}_3p_x)$ & 0.033 & $(\text{S}_2p_x,\text{S}_3p_x)$ & 0.043 \\

\end{tabular}
\end{table*}

Next, we discuss the band splitting of the conduction band. At the $K_\mathrm{c}$ point, the bottom of the conduction band in the monolayer and six-layers MoS$_2$ are mainly composed of Mo $d_{z^2}$, S $p_x$ and $p_y$ orbitals as shown in Fig.~\ref{Fig:weightband}. However, these orbitals contribute little to the band splitting due to the weak interlayer coupling as shown in Table~\ref{Table:Hamiltonian}. In contrast, we can see that the conduction band at the $Q_\mathrm{c}$ point has a large energy splitting as shown in Fig.~\ref{Fig:WANNIERlayerband}. This band splitting has been recognized that the overlap between S $p$ orbitals of neighboring layers plays a key role in strong interlayer interactions, particularly through $p_z$--$\,p_z$ coupling~\cite{Fang2015,Cappelluti2013}. However, we note that the interlayer $p_z$--$\,p_x$, and $p_z$--$\,p_y$ couplings have a non-negligible contribution to the band splitting at the $Q$ point as shown in Table~\ref{Table:Hamiltonian}. 

\begin{figure}
    \centering
    \includegraphics[width=0.9\linewidth]{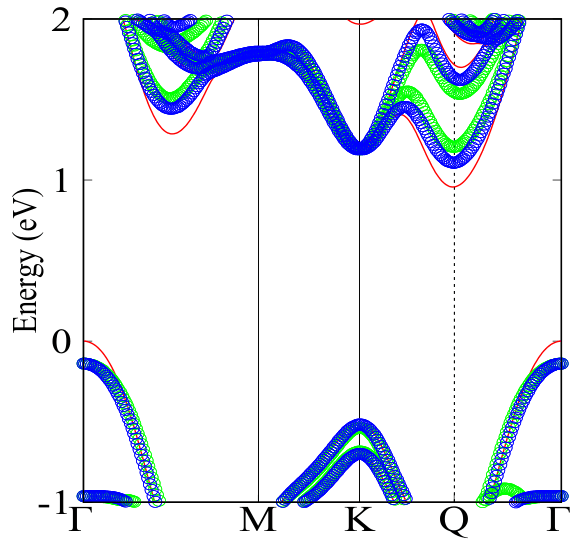}
    \caption{
    Band structures of bulk MoS$_2$ calculated using tight-binding (TB) models. The red line shows the result from the full TB Hamiltonian obtained using WANNIER90. The green and blue points show band structures from simplified TB Hamiltonians constructed by retaining selected interlayer couplings from this TB Hamiltonian. The green points include only S$_2$ $p_z$–S$_3$ $p_z$ coupling, while the blue points additionally include S$_2$ $p_z$–S$_3$ $p_x$, $p_y$ and S$_2$ $p_x$, $p_y$–S$_3$ $p_z$ couplings.
    }
    \label{fig:woband}
\end{figure}

Figure~\ref{fig:woband} shows the band structures obtained from several tight-binding (TB) models with different interlayer couplings. The red lines represent the full TB model constructed using Wannier90. The green points show the band structure obtained from a simplified TB Hamiltonian that retains only the interlayer coupling between the $p_z$ orbitals of sulfur atoms in neighboring layers. In this case, the conduction-band minimum at the $Q$ point is not reproduced. This indicates that the interlayer coupling between S $p_z$ orbitals alone is insufficient to capture the indirect $\Gamma$--$Q$ transition.
When the interlayer $p_z$--$p_x$ and $p_z$--$p_y$ couplings are additionally included (blue points), the conduction-band minimum shifts from the $K$ point to the $Q$ point. Therefore, to accurately describe the conduction-band minimum at the $Q$ point, it is necessary to include interlayer couplings involving not only the out-of-plane $p_z$ orbitals but also the in-plane $p_x$ and $p_y$ orbitals.

Finally, the above results can be visually grasped from the $\boldsymbol{k}$-resolved charge density $\rho_{n\boldsymbol{k}}$ which is defined by the following expression:
\begin{equation}
    \rho_{n\boldsymbol{k}}=\phi_{n\boldsymbol{k}}^*(\boldsymbol{r})\phi_{n\boldsymbol{k}}(\boldsymbol{r}),
\end{equation}
where $n$ is band index and $\phi_{n\boldsymbol{k}}(\boldsymbol{r})$ is the Kohn–Sham wavefunction. 
This $\boldsymbol{k}$-resolved charge density $\rho_{n\boldsymbol{k}}$ has the relationship with the charge density $\rho(\boldsymbol{r})$ as follows:
\begin{equation}
    \rho(\boldsymbol{r})=\sum_n \frac{1}{(2\pi)^3} \int d\boldsymbol{k} ~ \rho_{n \bm k}(\boldsymbol{r}).
\end{equation}

\begin{figure}[h]
\includegraphics[width=0.8\linewidth]{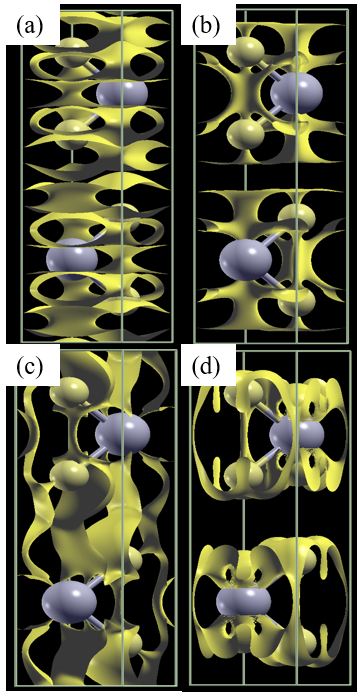}
\caption{The $\bm k$-resolved charge density distribution for bulk MoS$_2$ at (a) $\Gamma_\mathrm{v}$(b) $K_\mathrm{v}$ (c) $Q_\mathrm{c}$ (d) $K_\mathrm{c}$ point. These figures are drawn by XCrysDen\cite{xcrysden}.}
\label{Fig:K-cd}
\end{figure}

We computed the $\boldsymbol{k}$-resolved charge density $\rho_{n\boldsymbol{k}}$ using DFT calculation shown in Fig~\ref{Fig:QEWANNIERband} (a). Figures~\ref{Fig:K-cd}(a) and (c) show the charge density at the $\Gamma_\mathrm{v}$ and $Q_\mathrm{c}$ points, respectively. In both cases, the electron density is distributed between adjacent layers, indicating strong interlayer couplings. Figures~\ref{Fig:K-cd}(b) and (d) show the charge densities at $K_\mathrm{v}$ and $K_\mathrm{c}$ point. In contrast to Figs.~\ref{Fig:K-cd}(a) and (c), electrons are localized only within the layers with a small inter-layer distribution, indicating week interlayer couplings. The extent to which the charge density spreads into the interlayer region directly reflects the strength of the interlayer couplings. This provides a visual and intuitive picture of how the differences in interlayer coupling strength directly drive the band splitting.

\section{Conclusions}

In this study, we investigated the microscopic origin of the direct-indirect bandgap transition in MoS$_2$ induced by layer stacking, through atomic orbital analysis using the multilayer TB Hamiltonian of Eq.~\eqref{Hamiltonian}. From the viewpoint of the atomic orbitals, it has been recognized that the $p_z$--$\,p_z$ coupling between the sulfurs in neighboring layers plays an important role in the direct-indirect bandgap mechanism~\cite{Fang2015,Cappelluti2013}.
Our analysis based on the TB Hamiltonian~\eqref{Hamiltonian} demonstrated that S $p_z$ orbitals are the dominant factor in driving the band splitting at $\Gamma$ and $Q$ points while the $p_z$--$\,p_x$ $p_z$--$\,p_y$ couplings between the sulfur atoms in neighboring layers also make non-negligible contributions to the band splitting at the $Q_{\mathbf{c}}$ point. Indeed, we demonstrated that an accurate reproduction of the indirect $\Gamma$--$Q$ transition requires not only interlayer couplings between S $p$ orbitals in neighboring layers but also weak interlayer couplings between distant atomic sites. In addition, the $k$-resolved charge density provides an intuitive picture with clear physical meaning. Our results not only clarify the mechanism of the band splitting associated with the transition from a direct to an indirect band gap induced by the stacking of MoS$_2$ layers, but also suggest strategies for engineering its band structure through orbital and interlayer interaction control.

\section*{Acknowledgments}
 We are grateful to K. Tanno for the technical supports. We also thank M. Ochi for helpful comments and discussions.
This research is supported by JSPS KAKENHI Grants Numbers JP23H01130, JP24K00581, JP24K00588, JP25K00947, JP25K21684.

\clearpage

\bibliographystyle{apsrev4-1}
\bibliography{MoS2}

\begin{thebibliography}{25}%
\makeatletter
\providecommand \@ifxundefined [1]{%
 \@ifx{#1\undefined}
}%
\providecommand \@ifnum [1]{%
 \ifnum #1\expandafter \@firstoftwo
 \else \expandafter \@secondoftwo
 \fi
}%
\providecommand \@ifx [1]{%
 \ifx #1\expandafter \@firstoftwo
 \else \expandafter \@secondoftwo
 \fi
}%
\providecommand \natexlab [1]{#1}%
\providecommand \enquote  [1]{``#1''}%
\providecommand \bibnamefont  [1]{#1}%
\providecommand \bibfnamefont [1]{#1}%
\providecommand \citenamefont [1]{#1}%
\providecommand \href@noop [0]{\@secondoftwo}%
\providecommand \href [0]{\begingroup \@sanitize@url \@href}%
\providecommand \@href[1]{\@@startlink{#1}\@@href}%
\providecommand \@@href[1]{\endgroup#1\@@endlink}%
\providecommand \@sanitize@url [0]{\catcode `\\12\catcode `\$12\catcode `\&12\catcode `\#12\catcode `\^12\catcode `\_12\catcode `\%12\relax}%
\providecommand \@@startlink[1]{}%
\providecommand \@@endlink[0]{}%
\providecommand \url  [0]{\begingroup\@sanitize@url \@url }%
\providecommand \@url [1]{\endgroup\@href {#1}{\urlprefix }}%
\providecommand \urlprefix  [0]{URL }%
\providecommand \Eprint [0]{\href }%
\providecommand \doibase [0]{http://dx.doi.org/}%
\providecommand \selectlanguage [0]{\@gobble}%
\providecommand \bibinfo  [0]{\@secondoftwo}%
\providecommand \bibfield  [0]{\@secondoftwo}%
\providecommand \translation [1]{[#1]}%
\providecommand \BibitemOpen [0]{}%
\providecommand \bibitemStop [0]{}%
\providecommand \bibitemNoStop [0]{.\EOS\space}%
\providecommand \EOS [0]{\spacefactor3000\relax}%
\providecommand \BibitemShut  [1]{\csname bibitem#1\endcsname}%
\let\auto@bib@innerbib\@empty
\bibitem [{\citenamefont {Chhowalla}\ \emph {et~al.}(2015)\citenamefont {Chhowalla}, \citenamefont {Liu},\ and\ \citenamefont {Zhang}}]{chhowalla2015two}%
  \BibitemOpen
  \bibfield  {author} {\bibinfo {author} {\bibfnamefont {M.}~\bibnamefont {Chhowalla}}, \bibinfo {author} {\bibfnamefont {Z.}~\bibnamefont {Liu}}, \ and\ \bibinfo {author} {\bibfnamefont {H.}~\bibnamefont {Zhang}},\ }\href@noop {} {\bibfield  {journal} {\bibinfo  {journal} {Chem. Soc. Rev.}\ }\textbf {\bibinfo {volume} {44}},\ \bibinfo {pages} {2584} (\bibinfo {year} {2015})}\BibitemShut {NoStop}%
\bibitem [{\citenamefont {Liu}\ \emph {et~al.}(2015)\citenamefont {Liu}, \citenamefont {Xiao}, \citenamefont {Yao}, \citenamefont {Xu},\ and\ \citenamefont {Yao}}]{liu2015electronic}%
  \BibitemOpen
  \bibfield  {author} {\bibinfo {author} {\bibfnamefont {G.-B.}\ \bibnamefont {Liu}}, \bibinfo {author} {\bibfnamefont {D.}~\bibnamefont {Xiao}}, \bibinfo {author} {\bibfnamefont {Y.}~\bibnamefont {Yao}}, \bibinfo {author} {\bibfnamefont {X.}~\bibnamefont {Xu}}, \ and\ \bibinfo {author} {\bibfnamefont {W.}~\bibnamefont {Yao}},\ }\href@noop {} {\bibfield  {journal} {\bibinfo  {journal} {Chem. Soc. Rev.}\ }\textbf {\bibinfo {volume} {44}},\ \bibinfo {pages} {2643} (\bibinfo {year} {2015})}\BibitemShut {NoStop}%
\bibitem [{\citenamefont {Kuc}\ and\ \citenamefont {Heine}(2015)}]{kuc2015electronic}%
  \BibitemOpen
  \bibfield  {author} {\bibinfo {author} {\bibfnamefont {A.}~\bibnamefont {Kuc}}\ and\ \bibinfo {author} {\bibfnamefont {T.}~\bibnamefont {Heine}},\ }\href@noop {} {\bibfield  {journal} {\bibinfo  {journal} {Chem. Soc. Rev.}\ }\textbf {\bibinfo {volume} {44}},\ \bibinfo {pages} {2603} (\bibinfo {year} {2015})}\BibitemShut {NoStop}%
\bibitem [{\citenamefont {Li}\ and\ \citenamefont {Zhu}(2015)}]{LI2015}%
  \BibitemOpen
  \bibfield  {author} {\bibinfo {author} {\bibfnamefont {X.}~\bibnamefont {Li}}\ and\ \bibinfo {author} {\bibfnamefont {H.}~\bibnamefont {Zhu}},\ }\href@noop {} {\bibfield  {journal} {\bibinfo  {journal} {J. Materiomics}\ }\textbf {\bibinfo {volume} {1}},\ \bibinfo {pages} {33} (\bibinfo {year} {2015})}\BibitemShut {NoStop}%
\bibitem [{\citenamefont {Yazyev}\ and\ \citenamefont {Kis}(2015)}]{YAZYEV2015}%
  \BibitemOpen
  \bibfield  {author} {\bibinfo {author} {\bibfnamefont {O.~V.}\ \bibnamefont {Yazyev}}\ and\ \bibinfo {author} {\bibfnamefont {A.}~\bibnamefont {Kis}},\ }\href@noop {} {\bibfield  {journal} {\bibinfo  {journal} {Mater. Today}\ }\textbf {\bibinfo {volume} {18}},\ \bibinfo {pages} {20} (\bibinfo {year} {2015})}\BibitemShut {NoStop}%
\bibitem [{\citenamefont {Li}\ and\ \citenamefont {Galli}(2007)}]{Li2007}%
  \BibitemOpen
  \bibfield  {author} {\bibinfo {author} {\bibfnamefont {T.}~\bibnamefont {Li}}\ and\ \bibinfo {author} {\bibfnamefont {G.}~\bibnamefont {Galli}},\ }\href {\doibase 10.1021/jp075424v} {\bibfield  {journal} {\bibinfo  {journal} {J. Phys. Chem. C}\ }\textbf {\bibinfo {volume} {111}},\ \bibinfo {pages} {16192} (\bibinfo {year} {2007})}\BibitemShut {NoStop}%
\bibitem [{\citenamefont {Splendiani}\ \emph {et~al.}(2010)\citenamefont {Splendiani}, \citenamefont {Sun}, \citenamefont {Zhang}, \citenamefont {Li}, \citenamefont {Kim}, \citenamefont {Chim}, \citenamefont {Galli},\ and\ \citenamefont {Wang}}]{Splendiani2010}%
  \BibitemOpen
  \bibfield  {author} {\bibinfo {author} {\bibfnamefont {A.}~\bibnamefont {Splendiani}}, \bibinfo {author} {\bibfnamefont {L.}~\bibnamefont {Sun}}, \bibinfo {author} {\bibfnamefont {Y.}~\bibnamefont {Zhang}}, \bibinfo {author} {\bibfnamefont {T.}~\bibnamefont {Li}}, \bibinfo {author} {\bibfnamefont {J.}~\bibnamefont {Kim}}, \bibinfo {author} {\bibfnamefont {C.-Y.}\ \bibnamefont {Chim}}, \bibinfo {author} {\bibfnamefont {G.}~\bibnamefont {Galli}}, \ and\ \bibinfo {author} {\bibfnamefont {F.}~\bibnamefont {Wang}},\ }\href {\doibase 10.1021/nl903868w} {\bibfield  {journal} {\bibinfo  {journal} {Nano Lett.}\ }\textbf {\bibinfo {volume} {10}},\ \bibinfo {pages} {1271} (\bibinfo {year} {2010})}\BibitemShut {NoStop}%
\bibitem [{\citenamefont {Mak}\ \emph {et~al.}(2010)\citenamefont {Mak}, \citenamefont {Lee}, \citenamefont {Hone}, \citenamefont {Shan},\ and\ \citenamefont {Heinz}}]{MaK2010}%
  \BibitemOpen
  \bibfield  {author} {\bibinfo {author} {\bibfnamefont {K.~F.}\ \bibnamefont {Mak}}, \bibinfo {author} {\bibfnamefont {C.}~\bibnamefont {Lee}}, \bibinfo {author} {\bibfnamefont {J.}~\bibnamefont {Hone}}, \bibinfo {author} {\bibfnamefont {J.}~\bibnamefont {Shan}}, \ and\ \bibinfo {author} {\bibfnamefont {T.~F.}\ \bibnamefont {Heinz}},\ }\href@noop {} {\bibfield  {journal} {\bibinfo  {journal} {Phys. Rev. Lett.}\ }\textbf {\bibinfo {volume} {105}},\ \bibinfo {pages} {136805} (\bibinfo {year} {2010})}\BibitemShut {NoStop}%
\bibitem [{\citenamefont {Eda}\ \emph {et~al.}(2011)\citenamefont {Eda}, \citenamefont {Yamaguchi}, \citenamefont {Voiry}, \citenamefont {Fujita}, \citenamefont {Chen},\ and\ \citenamefont {Chhowalla}}]{Eda2011}%
  \BibitemOpen
  \bibfield  {author} {\bibinfo {author} {\bibfnamefont {G.}~\bibnamefont {Eda}}, \bibinfo {author} {\bibfnamefont {H.}~\bibnamefont {Yamaguchi}}, \bibinfo {author} {\bibfnamefont {D.}~\bibnamefont {Voiry}}, \bibinfo {author} {\bibfnamefont {T.}~\bibnamefont {Fujita}}, \bibinfo {author} {\bibfnamefont {M.}~\bibnamefont {Chen}}, \ and\ \bibinfo {author} {\bibfnamefont {M.}~\bibnamefont {Chhowalla}},\ }\href@noop {} {\bibfield  {journal} {\bibinfo  {journal} {Nano Lett.}\ }\textbf {\bibinfo {volume} {11}},\ \bibinfo {pages} {5111} (\bibinfo {year} {2011})}\BibitemShut {NoStop}%
\bibitem [{\citenamefont {Radisavljevic}\ \emph {et~al.}(2011)\citenamefont {Radisavljevic}, \citenamefont {Radenovic}, \citenamefont {Brivio}, \citenamefont {Giacometti},\ and\ \citenamefont {Kis}}]{Radisavljevic2011}%
  \BibitemOpen
  \bibfield  {author} {\bibinfo {author} {\bibfnamefont {B.}~\bibnamefont {Radisavljevic}}, \bibinfo {author} {\bibfnamefont {A.}~\bibnamefont {Radenovic}}, \bibinfo {author} {\bibfnamefont {J.}~\bibnamefont {Brivio}}, \bibinfo {author} {\bibfnamefont {V.}~\bibnamefont {Giacometti}}, \ and\ \bibinfo {author} {\bibfnamefont {A.}~\bibnamefont {Kis}},\ }\href {\doibase 10.1038/nnano.2010.279} {\bibfield  {journal} {\bibinfo  {journal} {Nat. Nanotechnol.}\ }\textbf {\bibinfo {volume} {6}},\ \bibinfo {pages} {147} (\bibinfo {year} {2011})}\BibitemShut {NoStop}%
\bibitem [{\citenamefont {Cappelluti}\ \emph {et~al.}(2013)\citenamefont {Cappelluti}, \citenamefont {Rold\'an}, \citenamefont {Silva-Guill\'en}, \citenamefont {Ordej\'on},\ and\ \citenamefont {Guinea}}]{Cappelluti2013}%
  \BibitemOpen
  \bibfield  {author} {\bibinfo {author} {\bibfnamefont {E.}~\bibnamefont {Cappelluti}}, \bibinfo {author} {\bibfnamefont {R.}~\bibnamefont {Rold\'an}}, \bibinfo {author} {\bibfnamefont {J.~A.}\ \bibnamefont {Silva-Guill\'en}}, \bibinfo {author} {\bibfnamefont {P.}~\bibnamefont {Ordej\'on}}, \ and\ \bibinfo {author} {\bibfnamefont {F.}~\bibnamefont {Guinea}},\ }\href {\doibase 10.1103/PhysRevB.88.075409} {\bibfield  {journal} {\bibinfo  {journal} {Phys. Rev. B}\ }\textbf {\bibinfo {volume} {88}},\ \bibinfo {pages} {075409} (\bibinfo {year} {2013})}\BibitemShut {NoStop}%
\bibitem [{\citenamefont {Zahid}\ \emph {et~al.}(2013)\citenamefont {Zahid}, \citenamefont {Liu}, \citenamefont {Zhu}, \citenamefont {Wang},\ and\ \citenamefont {Guo}}]{zahid2013}%
  \BibitemOpen
  \bibfield  {author} {\bibinfo {author} {\bibfnamefont {F.}~\bibnamefont {Zahid}}, \bibinfo {author} {\bibfnamefont {L.}~\bibnamefont {Liu}}, \bibinfo {author} {\bibfnamefont {Y.}~\bibnamefont {Zhu}}, \bibinfo {author} {\bibfnamefont {J.}~\bibnamefont {Wang}}, \ and\ \bibinfo {author} {\bibfnamefont {H.}~\bibnamefont {Guo}},\ }\href@noop {} {\bibfield  {journal} {\bibinfo  {journal} {Aip Advances}\ }\textbf {\bibinfo {volume} {3}} (\bibinfo {year} {2013})}\BibitemShut {NoStop}%
\bibitem [{\citenamefont {Liu}\ \emph {et~al.}(2013)\citenamefont {Liu}, \citenamefont {Shan}, \citenamefont {Yao}, \citenamefont {Yao},\ and\ \citenamefont {Xiao}}]{liu2013}%
  \BibitemOpen
  \bibfield  {author} {\bibinfo {author} {\bibfnamefont {G.-B.}\ \bibnamefont {Liu}}, \bibinfo {author} {\bibfnamefont {W.-Y.}\ \bibnamefont {Shan}}, \bibinfo {author} {\bibfnamefont {Y.}~\bibnamefont {Yao}}, \bibinfo {author} {\bibfnamefont {W.}~\bibnamefont {Yao}}, \ and\ \bibinfo {author} {\bibfnamefont {D.}~\bibnamefont {Xiao}},\ }\href@noop {} {\bibfield  {journal} {\bibinfo  {journal} {Physical Review B―Condensed Matter and Materials Physics}\ }\textbf {\bibinfo {volume} {88}},\ \bibinfo {pages} {085433} (\bibinfo {year} {2013})}\BibitemShut {NoStop}%
\bibitem [{\citenamefont {Rostami}\ \emph {et~al.}(2013)\citenamefont {Rostami}, \citenamefont {Moghaddam},\ and\ \citenamefont {Asgari}}]{Rostami2013}%
  \BibitemOpen
  \bibfield  {author} {\bibinfo {author} {\bibfnamefont {H.}~\bibnamefont {Rostami}}, \bibinfo {author} {\bibfnamefont {A.~G.}\ \bibnamefont {Moghaddam}}, \ and\ \bibinfo {author} {\bibfnamefont {R.}~\bibnamefont {Asgari}},\ }\href {\doibase 10.1103/PhysRevB.88.085440} {\bibfield  {journal} {\bibinfo  {journal} {Phys. Rev. B}\ }\textbf {\bibinfo {volume} {88}},\ \bibinfo {pages} {085440} (\bibinfo {year} {2013})}\BibitemShut {NoStop}%
\bibitem [{\citenamefont {Ridolfi}\ \emph {et~al.}(2015)\citenamefont {Ridolfi}, \citenamefont {Le}, \citenamefont {Rahman}, \citenamefont {Mucciolo},\ and\ \citenamefont {Lewenkopf}}]{ridolfi2015}%
  \BibitemOpen
  \bibfield  {author} {\bibinfo {author} {\bibfnamefont {E.}~\bibnamefont {Ridolfi}}, \bibinfo {author} {\bibfnamefont {D.}~\bibnamefont {Le}}, \bibinfo {author} {\bibfnamefont {T.~S.}\ \bibnamefont {Rahman}}, \bibinfo {author} {\bibfnamefont {E.~R.}\ \bibnamefont {Mucciolo}}, \ and\ \bibinfo {author} {\bibfnamefont {C.~H.}\ \bibnamefont {Lewenkopf}},\ }\href@noop {} {\bibfield  {journal} {\bibinfo  {journal} {J. Phys. Condens. Matter}\ }\textbf {\bibinfo {volume} {27}},\ \bibinfo {pages} {365501} (\bibinfo {year} {2015})}\BibitemShut {NoStop}%
\bibitem [{\citenamefont {Fang}\ \emph {et~al.}(2015)\citenamefont {Fang}, \citenamefont {Kuate~Defo}, \citenamefont {Shirodkar}, \citenamefont {Lieu}, \citenamefont {Tritsaris},\ and\ \citenamefont {Kaxiras}}]{Fang2015}%
  \BibitemOpen
  \bibfield  {author} {\bibinfo {author} {\bibfnamefont {S.}~\bibnamefont {Fang}}, \bibinfo {author} {\bibfnamefont {R.}~\bibnamefont {Kuate~Defo}}, \bibinfo {author} {\bibfnamefont {S.~N.}\ \bibnamefont {Shirodkar}}, \bibinfo {author} {\bibfnamefont {S.}~\bibnamefont {Lieu}}, \bibinfo {author} {\bibfnamefont {G.~A.}\ \bibnamefont {Tritsaris}}, \ and\ \bibinfo {author} {\bibfnamefont {E.}~\bibnamefont {Kaxiras}},\ }\href {\doibase 10.1103/PhysRevB.92.205108} {\bibfield  {journal} {\bibinfo  {journal} {Phys. Rev. B}\ }\textbf {\bibinfo {volume} {92}},\ \bibinfo {pages} {205108} (\bibinfo {year} {2015})}\BibitemShut {NoStop}%
\bibitem [{\citenamefont {Dickinson}\ and\ \citenamefont {Pauling}(1923)}]{Pauling1923}%
  \BibitemOpen
  \bibfield  {author} {\bibinfo {author} {\bibfnamefont {R.~G.}\ \bibnamefont {Dickinson}}\ and\ \bibinfo {author} {\bibfnamefont {L.}~\bibnamefont {Pauling}},\ }\href@noop {} {\bibfield  {journal} {\bibinfo  {journal} {J. Am. Chem. Soc.}\ }\textbf {\bibinfo {volume} {45}},\ \bibinfo {pages} {1466} (\bibinfo {year} {1923})}\BibitemShut {NoStop}%
\bibitem [{\citenamefont {Momma}\ and\ \citenamefont {Izumi}(2011)}]{VESTA}%
  \BibitemOpen
  \bibfield  {author} {\bibinfo {author} {\bibfnamefont {K.}~\bibnamefont {Momma}}\ and\ \bibinfo {author} {\bibfnamefont {F.}~\bibnamefont {Izumi}},\ }\href@noop {} {\bibfield  {journal} {\bibinfo  {journal} {J. Appl. Cryst.}\ }\textbf {\bibinfo {volume} {44}},\ \bibinfo {pages} {1272} (\bibinfo {year} {2011})}\BibitemShut {NoStop}%
\bibitem [{\citenamefont {Giannozzi}\ \emph {et~al.}(2017)\citenamefont {Giannozzi}, \citenamefont {Andreussi}, \citenamefont {Brumme}, \citenamefont {Bunau}, \citenamefont {Buongiorno~Nardelli}, \citenamefont {Calandra}, \citenamefont {Car}, \citenamefont {Cavazzoni}, \citenamefont {Ceresoli}, \citenamefont {Cococcioni}, \citenamefont {Colonna}, \citenamefont {Carnimeo}, \citenamefont {Dal~Corso}, \citenamefont {de~Gironcoli}, \citenamefont {Delugas}, \citenamefont {DiStasio}, \citenamefont {Ferretti}, \citenamefont {Floris}, \citenamefont {Fratesi}, \citenamefont {Fugallo}, \citenamefont {Gebauer}, \citenamefont {Gerstmann}, \citenamefont {Giustino}, \citenamefont {Gorni}, \citenamefont {Jia}, \citenamefont {Kawamura}, \citenamefont {Ko}, \citenamefont {Kokalj}, \citenamefont {K^^c3^^bc^^c3^^a7^^c3^^bckbenli}, \citenamefont {Lazzeri}, \citenamefont {Marsili}, \citenamefont {Marzari}, \citenamefont {Mauri}, \citenamefont {Nguyen}, \citenamefont {Nguyen}, \citenamefont {Otero-de-la Roza}, \citenamefont
  {Paulatto}, \citenamefont {Ponc^^c3^^a9}, \citenamefont {Rocca}, \citenamefont {Sabatini}, \citenamefont {Santra}, \citenamefont {Schlipf}, \citenamefont {Seitsonen}, \citenamefont {Smogunov}, \citenamefont {Timrov}, \citenamefont {Thonhauser}, \citenamefont {Umari}, \citenamefont {Vast}, \citenamefont {Wu},\ and\ \citenamefont {Baroni}}]{QE1}%
  \BibitemOpen
  \bibfield  {author} {\bibinfo {author} {\bibfnamefont {P.}~\bibnamefont {Giannozzi}}, \bibinfo {author} {\bibfnamefont {O.}~\bibnamefont {Andreussi}}, \bibinfo {author} {\bibfnamefont {T.}~\bibnamefont {Brumme}}, \bibinfo {author} {\bibfnamefont {O.}~\bibnamefont {Bunau}}, \bibinfo {author} {\bibfnamefont {M.}~\bibnamefont {Buongiorno~Nardelli}}, \bibinfo {author} {\bibfnamefont {M.}~\bibnamefont {Calandra}}, \bibinfo {author} {\bibfnamefont {R.}~\bibnamefont {Car}}, \bibinfo {author} {\bibfnamefont {C.}~\bibnamefont {Cavazzoni}}, \bibinfo {author} {\bibfnamefont {D.}~\bibnamefont {Ceresoli}}, \bibinfo {author} {\bibfnamefont {M.}~\bibnamefont {Cococcioni}}, \bibinfo {author} {\bibfnamefont {N.}~\bibnamefont {Colonna}}, \bibinfo {author} {\bibfnamefont {I.}~\bibnamefont {Carnimeo}}, \bibinfo {author} {\bibfnamefont {A.}~\bibnamefont {Dal~Corso}}, \bibinfo {author} {\bibfnamefont {S.}~\bibnamefont {de~Gironcoli}}, \bibinfo {author} {\bibfnamefont {P.}~\bibnamefont {Delugas}}, \bibinfo {author} {\bibfnamefont
  {R.~A.}\ \bibnamefont {DiStasio}}, \bibinfo {author} {\bibfnamefont {A.}~\bibnamefont {Ferretti}}, \bibinfo {author} {\bibfnamefont {A.}~\bibnamefont {Floris}}, \bibinfo {author} {\bibfnamefont {G.}~\bibnamefont {Fratesi}}, \bibinfo {author} {\bibfnamefont {G.}~\bibnamefont {Fugallo}}, \bibinfo {author} {\bibfnamefont {R.}~\bibnamefont {Gebauer}}, \bibinfo {author} {\bibfnamefont {U.}~\bibnamefont {Gerstmann}}, \bibinfo {author} {\bibfnamefont {F.}~\bibnamefont {Giustino}}, \bibinfo {author} {\bibfnamefont {T.}~\bibnamefont {Gorni}}, \bibinfo {author} {\bibfnamefont {J.}~\bibnamefont {Jia}}, \bibinfo {author} {\bibfnamefont {M.}~\bibnamefont {Kawamura}}, \bibinfo {author} {\bibfnamefont {H.-Y.}\ \bibnamefont {Ko}}, \bibinfo {author} {\bibfnamefont {A.}~\bibnamefont {Kokalj}}, \bibinfo {author} {\bibfnamefont {E.}~\bibnamefont {K^^c3^^bc^^c3^^a7^^c3^^bckbenli}}, \bibinfo {author} {\bibfnamefont {M.}~\bibnamefont {Lazzeri}}, \bibinfo {author} {\bibfnamefont {M.}~\bibnamefont {Marsili}}, \bibinfo {author}
  {\bibfnamefont {N.}~\bibnamefont {Marzari}}, \bibinfo {author} {\bibfnamefont {F.}~\bibnamefont {Mauri}}, \bibinfo {author} {\bibfnamefont {N.~L.}\ \bibnamefont {Nguyen}}, \bibinfo {author} {\bibfnamefont {H.-V.}\ \bibnamefont {Nguyen}}, \bibinfo {author} {\bibfnamefont {A.}~\bibnamefont {Otero-de-la Roza}}, \bibinfo {author} {\bibfnamefont {L.}~\bibnamefont {Paulatto}}, \bibinfo {author} {\bibfnamefont {S.}~\bibnamefont {Ponc^^c3^^a9}}, \bibinfo {author} {\bibfnamefont {D.}~\bibnamefont {Rocca}}, \bibinfo {author} {\bibfnamefont {R.}~\bibnamefont {Sabatini}}, \bibinfo {author} {\bibfnamefont {B.}~\bibnamefont {Santra}}, \bibinfo {author} {\bibfnamefont {M.}~\bibnamefont {Schlipf}}, \bibinfo {author} {\bibfnamefont {A.~P.}\ \bibnamefont {Seitsonen}}, \bibinfo {author} {\bibfnamefont {A.}~\bibnamefont {Smogunov}}, \bibinfo {author} {\bibfnamefont {I.}~\bibnamefont {Timrov}}, \bibinfo {author} {\bibfnamefont {T.}~\bibnamefont {Thonhauser}}, \bibinfo {author} {\bibfnamefont {P.}~\bibnamefont {Umari}}, \bibinfo
  {author} {\bibfnamefont {N.}~\bibnamefont {Vast}}, \bibinfo {author} {\bibfnamefont {X.}~\bibnamefont {Wu}}, \ and\ \bibinfo {author} {\bibfnamefont {S.}~\bibnamefont {Baroni}},\ }\href@noop {} {\bibfield  {journal} {\bibinfo  {journal} {J. Phys. Condens. Matter}\ }\textbf {\bibinfo {volume} {29}},\ \bibinfo {pages} {465901} (\bibinfo {year} {2017})}\BibitemShut {NoStop}%
\bibitem [{\citenamefont {Giannozzi}\ \emph {et~al.}(2009)\citenamefont {Giannozzi}, \citenamefont {Baroni}, \citenamefont {Bonini}, \citenamefont {Calandra}, \citenamefont {Car}, \citenamefont {Cavazzoni}, \citenamefont {Ceresoli}, \citenamefont {Chiarotti}, \citenamefont {Cococcioni}, \citenamefont {Dabo}, \citenamefont {Dal~Corso}, \citenamefont {de~Gironcoli}, \citenamefont {Fabris}, \citenamefont {Fratesi}, \citenamefont {Gebauer}, \citenamefont {Gerstmann}, \citenamefont {Gougoussis}, \citenamefont {Kokalj}, \citenamefont {Lazzeri}, \citenamefont {Martin-Samos}, \citenamefont {Marzari}, \citenamefont {Mauri}, \citenamefont {Mazzarello}, \citenamefont {Paolini}, \citenamefont {Pasquarello}, \citenamefont {Paulatto}, \citenamefont {Sbraccia}, \citenamefont {Scandolo}, \citenamefont {Sclauzero}, \citenamefont {Seitsonen}, \citenamefont {Smogunov}, \citenamefont {Umari},\ and\ \citenamefont {Wentzcovitch}}]{QE2}%
  \BibitemOpen
  \bibfield  {author} {\bibinfo {author} {\bibfnamefont {P.}~\bibnamefont {Giannozzi}}, \bibinfo {author} {\bibfnamefont {S.}~\bibnamefont {Baroni}}, \bibinfo {author} {\bibfnamefont {N.}~\bibnamefont {Bonini}}, \bibinfo {author} {\bibfnamefont {M.}~\bibnamefont {Calandra}}, \bibinfo {author} {\bibfnamefont {R.}~\bibnamefont {Car}}, \bibinfo {author} {\bibfnamefont {C.}~\bibnamefont {Cavazzoni}}, \bibinfo {author} {\bibfnamefont {D.}~\bibnamefont {Ceresoli}}, \bibinfo {author} {\bibfnamefont {G.~L.}\ \bibnamefont {Chiarotti}}, \bibinfo {author} {\bibfnamefont {M.}~\bibnamefont {Cococcioni}}, \bibinfo {author} {\bibfnamefont {I.}~\bibnamefont {Dabo}}, \bibinfo {author} {\bibfnamefont {A.}~\bibnamefont {Dal~Corso}}, \bibinfo {author} {\bibfnamefont {S.}~\bibnamefont {de~Gironcoli}}, \bibinfo {author} {\bibfnamefont {S.}~\bibnamefont {Fabris}}, \bibinfo {author} {\bibfnamefont {G.}~\bibnamefont {Fratesi}}, \bibinfo {author} {\bibfnamefont {R.}~\bibnamefont {Gebauer}}, \bibinfo {author} {\bibfnamefont
  {U.}~\bibnamefont {Gerstmann}}, \bibinfo {author} {\bibfnamefont {C.}~\bibnamefont {Gougoussis}}, \bibinfo {author} {\bibfnamefont {A.}~\bibnamefont {Kokalj}}, \bibinfo {author} {\bibfnamefont {M.}~\bibnamefont {Lazzeri}}, \bibinfo {author} {\bibfnamefont {L.}~\bibnamefont {Martin-Samos}}, \bibinfo {author} {\bibfnamefont {N.}~\bibnamefont {Marzari}}, \bibinfo {author} {\bibfnamefont {F.}~\bibnamefont {Mauri}}, \bibinfo {author} {\bibfnamefont {R.}~\bibnamefont {Mazzarello}}, \bibinfo {author} {\bibfnamefont {S.}~\bibnamefont {Paolini}}, \bibinfo {author} {\bibfnamefont {A.}~\bibnamefont {Pasquarello}}, \bibinfo {author} {\bibfnamefont {L.}~\bibnamefont {Paulatto}}, \bibinfo {author} {\bibfnamefont {C.}~\bibnamefont {Sbraccia}}, \bibinfo {author} {\bibfnamefont {S.}~\bibnamefont {Scandolo}}, \bibinfo {author} {\bibfnamefont {G.}~\bibnamefont {Sclauzero}}, \bibinfo {author} {\bibfnamefont {A.~P.}\ \bibnamefont {Seitsonen}}, \bibinfo {author} {\bibfnamefont {A.}~\bibnamefont {Smogunov}}, \bibinfo {author}
  {\bibfnamefont {P.}~\bibnamefont {Umari}}, \ and\ \bibinfo {author} {\bibfnamefont {R.~M.}\ \bibnamefont {Wentzcovitch}},\ }\href@noop {} {\bibfield  {journal} {\bibinfo  {journal} {J. Phys. Condens. Matter}\ }\textbf {\bibinfo {volume} {21}},\ \bibinfo {pages} {395502} (\bibinfo {year} {2009})}\BibitemShut {NoStop}%
\bibitem [{\citenamefont {Pizzi}\ \emph {et~al.}(2020)\citenamefont {Pizzi}, \citenamefont {Vitale}, \citenamefont {Arita}, \citenamefont {Bl^^c3^^bcgel}, \citenamefont {Freimuth}, \citenamefont {G^^c3^^a9ranton}, \citenamefont {Gibertini}, \citenamefont {Gresch}, \citenamefont {Johnson}, \citenamefont {Koretsune}, \citenamefont {Iba^^c3^^b1ez-Azpiroz}, \citenamefont {Lee}, \citenamefont {Lihm}, \citenamefont {Marchand}, \citenamefont {Marrazzo}, \citenamefont {Mokrousov}, \citenamefont {Mustafa}, \citenamefont {Nohara}, \citenamefont {Nomura}, \citenamefont {Paulatto}, \citenamefont {Ponc^^c3^^a9}, \citenamefont {Ponweiser}, \citenamefont {Qiao}, \citenamefont {Th^^c3^^b6le}, \citenamefont {Tsirkin}, \citenamefont {Wierzbowska}, \citenamefont {Marzari}, \citenamefont {Vanderbilt}, \citenamefont {Souza}, \citenamefont {Mostofi},\ and\ \citenamefont {Yates}}]{WANNIER}%
  \BibitemOpen
  \bibfield  {author} {\bibinfo {author} {\bibfnamefont {G.}~\bibnamefont {Pizzi}}, \bibinfo {author} {\bibfnamefont {V.}~\bibnamefont {Vitale}}, \bibinfo {author} {\bibfnamefont {R.}~\bibnamefont {Arita}}, \bibinfo {author} {\bibfnamefont {S.}~\bibnamefont {Bl^^c3^^bcgel}}, \bibinfo {author} {\bibfnamefont {F.}~\bibnamefont {Freimuth}}, \bibinfo {author} {\bibfnamefont {G.}~\bibnamefont {G^^c3^^a9ranton}}, \bibinfo {author} {\bibfnamefont {M.}~\bibnamefont {Gibertini}}, \bibinfo {author} {\bibfnamefont {D.}~\bibnamefont {Gresch}}, \bibinfo {author} {\bibfnamefont {C.}~\bibnamefont {Johnson}}, \bibinfo {author} {\bibfnamefont {T.}~\bibnamefont {Koretsune}}, \bibinfo {author} {\bibfnamefont {J.}~\bibnamefont {Iba^^c3^^b1ez-Azpiroz}}, \bibinfo {author} {\bibfnamefont {H.}~\bibnamefont {Lee}}, \bibinfo {author} {\bibfnamefont {J.-M.}\ \bibnamefont {Lihm}}, \bibinfo {author} {\bibfnamefont {D.}~\bibnamefont {Marchand}}, \bibinfo {author} {\bibfnamefont {A.}~\bibnamefont {Marrazzo}}, \bibinfo {author}
  {\bibfnamefont {Y.}~\bibnamefont {Mokrousov}}, \bibinfo {author} {\bibfnamefont {J.~I.}\ \bibnamefont {Mustafa}}, \bibinfo {author} {\bibfnamefont {Y.}~\bibnamefont {Nohara}}, \bibinfo {author} {\bibfnamefont {Y.}~\bibnamefont {Nomura}}, \bibinfo {author} {\bibfnamefont {L.}~\bibnamefont {Paulatto}}, \bibinfo {author} {\bibfnamefont {S.}~\bibnamefont {Ponc^^c3^^a9}}, \bibinfo {author} {\bibfnamefont {T.}~\bibnamefont {Ponweiser}}, \bibinfo {author} {\bibfnamefont {J.}~\bibnamefont {Qiao}}, \bibinfo {author} {\bibfnamefont {F.}~\bibnamefont {Th^^c3^^b6le}}, \bibinfo {author} {\bibfnamefont {S.~S.}\ \bibnamefont {Tsirkin}}, \bibinfo {author} {\bibfnamefont {M.}~\bibnamefont {Wierzbowska}}, \bibinfo {author} {\bibfnamefont {N.}~\bibnamefont {Marzari}}, \bibinfo {author} {\bibfnamefont {D.}~\bibnamefont {Vanderbilt}}, \bibinfo {author} {\bibfnamefont {I.}~\bibnamefont {Souza}}, \bibinfo {author} {\bibfnamefont {A.~A.}\ \bibnamefont {Mostofi}}, \ and\ \bibinfo {author} {\bibfnamefont {J.~R.}\ \bibnamefont
  {Yates}},\ }\href@noop {} {\bibfield  {journal} {\bibinfo  {journal} {J. Phys. Condens. Matter}\ }\textbf {\bibinfo {volume} {32}},\ \bibinfo {pages} {165902} (\bibinfo {year} {2020})}\BibitemShut {NoStop}%
\bibitem [{\citenamefont {Cheiwchanchamnangij}\ and\ \citenamefont {Lambrecht}(2012)}]{cheiw2012}%
  \BibitemOpen
  \bibfield  {author} {\bibinfo {author} {\bibfnamefont {T.}~\bibnamefont {Cheiwchanchamnangij}}\ and\ \bibinfo {author} {\bibfnamefont {W.~R.}\ \bibnamefont {Lambrecht}},\ }\href@noop {} {\bibfield  {journal} {\bibinfo  {journal} {Phys. Rev. B}\ }\textbf {\bibinfo {volume} {85}},\ \bibinfo {pages} {205302} (\bibinfo {year} {2012})}\BibitemShut {NoStop}%
\bibitem [{\citenamefont {Kadantsev}\ and\ \citenamefont {Hawrylak}(2012)}]{EugeneS2012}%
  \BibitemOpen
  \bibfield  {author} {\bibinfo {author} {\bibfnamefont {E.~S.}\ \bibnamefont {Kadantsev}}\ and\ \bibinfo {author} {\bibfnamefont {P.}~\bibnamefont {Hawrylak}},\ }\href@noop {} {\bibfield  {journal} {\bibinfo  {journal} {Solid State Commun.}\ }\textbf {\bibinfo {volume} {152}},\ \bibinfo {pages} {909} (\bibinfo {year} {2012})}\BibitemShut {NoStop}%
\bibitem [{\citenamefont {Ahmad}\ and\ \citenamefont {Mukherjee}(2014)}]{ahmad2014comparative}%
  \BibitemOpen
  \bibfield  {author} {\bibinfo {author} {\bibfnamefont {S.}~\bibnamefont {Ahmad}}\ and\ \bibinfo {author} {\bibfnamefont {S.}~\bibnamefont {Mukherjee}},\ }\href@noop {} {\bibfield  {journal} {\bibinfo  {journal} {Graphene}\ }\textbf {\bibinfo {volume} {3}},\ \bibinfo {pages} {52} (\bibinfo {year} {2014})}\BibitemShut {NoStop}%
\bibitem [{\citenamefont {Kokalj}(1999)}]{xcrysden}%
  \BibitemOpen
  \bibfield  {author} {\bibinfo {author} {\bibfnamefont {A.}~\bibnamefont {Kokalj}},\ }\href@noop {} {\bibfield  {journal} {\bibinfo  {journal} {J. Mol. Graph. Model.}\ }\textbf {\bibinfo {volume} {17}},\ \bibinfo {pages} {176} (\bibinfo {year} {1999})}\BibitemShut {NoStop}%
\end{thebibliography}%

\end{document}